# Non-destructive Imaging of Individual Bio-Molecules

Matthias Germann, Tatiana Latychevskaia, Conrad Escher & Hans-Werner Fink

Institute of Physics, University of Zurich, Winterthurerstrasse 190, CH-8057 Zurich, Switzerland

61.80.-x, 61.05.jp, 42.40.-i, 87.14.gk

**Radiation damage is considered to be the major problem that still prevents imaging an individual biological molecule for structural analysis. So far, all known mapping techniques using sufficient short wave-length radiation, be it X-rays or high energy electrons, circumvent this problem by averaging over many molecules. Averaging, however, leaves conformational details uncovered. Even the anticipated use of ultra-short but extremely bright X-ray bursts of a Free Electron Laser shall afford averaging over $10^6$ molecules to arrive at atomic resolution. Here we present direct experimental evidence for non-destructive imaging of individual DNA molecules. In fact, we show that DNA withstands coherent low energy electron radiation with deBroglie wavelength in the Ångstrom regime despite a vast dose of $10^8$ electrons/nm$^2$ accumulated over more than one hour.**

Exploring the three-dimensional structure of individual biomolecules, in particular those of proteins, is the foundation for a basic understanding of bio-chemistry, molecular biology and bio-physics. Most of the protein structural information data available today have been obtained from crystallography experiments by averaging over many molecules assembled into a crystal. Despite this vast amount of available data, a strong desire for acquiring structural data from just one individual molecule is emerging for good reasons. Most of the relevant biological molecules exhibit different conformations; thus averaging does not reveal detailed structural information. Moreover, there is large quantity of proteins, in particular the important class of membrane proteins, featuring a pronounced reluctance to readily crystallize.

Due to the strong inelastic scattering of X-rays and high energy electrons there is little hope for obtaining structural information from just one single molecule by conventional X-ray or high energy electron microscopy tools. Despite recent advances in cryo-electron microscopy, especially in image processing and reconstruction, averaging over typically 10,000 images is still necessary to build up a high signal to noise ratio image with structural features finally emerging[1]. This in turn smears out most of the details related to conformational flexibility. The necessity for averaging is given by radiation damage inherent to the interaction with high energy electrons and limits the obtainable resolution in conventional electron microscopy to ~1nanometer[2]. In order to obtain structures of individual biological molecules at atomic resolution, new concepts and technologies are envisioned. A major effort currently underway involves the development and implementation of the X-ray Free Electron Laser (XFEL), as a source of ultra short but extremely intense X-ray pulses. The overall idea is to take advantage of the principle of inertia: by keeping the interaction time of the intense X-ray burst with the molecule of interest extremely short, the site information of the atoms is carried to the detector before the molecule has been given time to finally decompose[3]. Unfortunately, in X-ray diffraction, inelastic scattering outweighs elastic scattering but only the latter carries information about the structure of molecules. Hence, a very large number, of the order of $10^6$, diffraction patterns of identical molecules must be recorded in order to obtain structural detail at a resolution of 3Å, even with a 10fs X-ray pulse containing $2x10^{12}$ photons at 1.5Å wave length[4].

In the following we show that a molecule as fragile as DNA withstands irradiation by coherent low energy electrons and remains unperturbed even after a total dose of at least 5 orders of magnitude larger than the permissible dose in X-ray or high energy electron imaging. The experimental set-up for testing radiation damage is illustrated in Fig. 1. DNA molecules are stretched over holes by using freeze drying technology known from cryo-microscopy[5]. First, an array of 1μm diameter holes in a thin carbon film is cleaned and rendered hydrophilic by UV-ozone treatment. Next, a droplet of λ-DNA solution of 2μg/ml concentration is applied onto the carbon film. Following an incubation time of typically 10 minutes, blotting paper is used to remove excess water. The remaining thin water film is transformed into amorphous ice by rapid quenching in liquid ethane. Next, the sample is freeze-dried at -80°C under vacuum conditions within typically 20 minutes and monitored by a mass-spectrometer above the sample. Finally, the sample is transferred into the Low Energy Electron Point Source (LEEPS)-microscope chamber for obtaining DNA electron holograms[6] at various energies ranging from a few 10eV up to about 300eV.

For the holographic imaging solely the elastic scattered electrons contribute to the hologram, while the inelastic scattered electrons lead to an incoherent diffuse background at the detector level. The fraction of inelastically scattered electrons by an object can be estimated by negatively biasing the front of the detector. While there is a significant amount of inelastic scattering in imaging metals with low energy electrons, measurable inelastic scattering in imaging DNA has not been observed. This could be a first, albeit just qualitative, hint for little or no radiation damage caused by low energy electrons. To actually measure the electron dose leaving DNA molecules unperturbed, we have carried out quantitative experiments. Evidence for damage-free imaging is provided in Fig. 2 showing a hologram of DNA molecules subject to 60eV electron radiation. In this experiment DNA has continuously been exposed to a 200nA electron current for 70 minutes. A set of DNA holograms has been recorded every 10 minutes. Next, three regions in the DNA holograms have been selected, marked by red squares in Fig. 2. Thereafter, the cross-correlation function between the first holographic record and subsequent holograms taken at 10min intervals have been computed for these very regions.

As evident from Fig. 2, the cross-correlation coefficient varies between 0.93 and 1.0 indicating a high degree of similarity between the first and all subsequent holograms. While the fluctuation of the cross-correlation coefficient is apparent, its time dependence differs in all three DNA hologram regions. These fluctuations are due to statistical noise which varies from hologram to hologram and is attributed to the intrinsic stochastic process of field emission and to detector noise. As a consequence, the cross-correlation coefficient reaches values just below unity at best. However, there is no tendency of the cross-correlation coefficient to decay in time. In fact, the coefficient persists above 0.93 even after 70 minutes of continuous exposure. Thus, the molecule's structure remained intact during 70 minutes of exposure to 60 eV electrons. The total accumulated dose during that time amounts to $10^8$ electrons/nm$^2$. This remarkably high electron dose certainly provides enough scattering events in a single molecule to be able to extract structural information at Ångstrom resolution. Moreover, there is nothing to be said against increasing the imaging current from 200nA into the μA range or prolonging the exposure time. The kinetic energy of 60eV has been chosen here because the corresponding deBroglie wave length is close to the 1.5Å X-ray wave length used in XFEL simulations setting the boundaries for single molecule imaging. However, similar experiments as described above have also been carried out at 110eV electron energies and quantitatively analyzed revealing the same result of no observable damage to DNA. To demonstrate that our findings depend only on electron energy but not on any particularities of our set-up, we would now like to present a control experiment, done in the same manner but at higher electron energies of 260eV where DNA actually decomposes rapidly within a few seconds.

In fact, a detailed analysis as in the non-damaging situation described above has not been possible here because of a too rapid decomposition of the molecules within the first 10 seconds of the observation process. The situation is illustrated in Fig. 3. Again, DNA is stretched over a 1μm diameter hole in a carbon film but now imaged with 260eV energy electrons. Even before completing the data acquisition for the first image by the slow-scan CCD camera, the molecule had already been partly damaged as apparent in Fig. 3(a). Evidently the 260eV electrons cause bond-braking in DNA. Since the DNA does not rest on a support but is free-standing, small molecular fragments created while DNA is decomposing sublimate into the vacuum. As a consequence, the remaining DNA gets shorter and shorter within a few seconds of observation.

For comparison to the non-damaging case, the evolution of the cross-correlation coefficient has also been computed and is shown in Fig. 3(e). As expected, the process of DNA decomposition is accompanied by a rapid decrease of the cross correlation coefficient between subsequent DNA images. The red bar displayed at the top of Fig. 3(e) illustrates the range of fluctuations of the cross-correlation coefficient in the non-damaging experiment. While 60eV electrons apparently cause no detectable damage to DNA at all, a rapid decomposition on a 4000 times shorter time scale is observed when DNA is subject to the interaction with 260eV electrons. We would like to point out that 60eV kinetic energy of the imaging electron wave is not a unique value or condition for gently imaging DNA molecules. Equally non-destructive imaging conditions have empirically been found also at 115, 140, 215 and 230eV kinetic electron energy for example.

Apparently, there are low electron energy ranges, where DNA experiences rapid damage. However, fortunately there are also regimes where DNA can be imaged using an extremely high dose without any damage at all. The permissible dose leaving a molecule unperturbed is at least 5 orders of magnitude greater than in conventional X-rays or high energy electrons imaging, demonstrating that coherent low energy electrons are the only non-damaging Ångstrom wave lengths radiation. With coherent low energy electrons it shall thus be possible to look at truly just one entity if it comes to high resolution diffraction microscopy of individual bio molecules.

## Acknowledgements

The work presented here is supported by the Project SIBMAR, part of the "New and Emerging Science and Technology" European Programme.

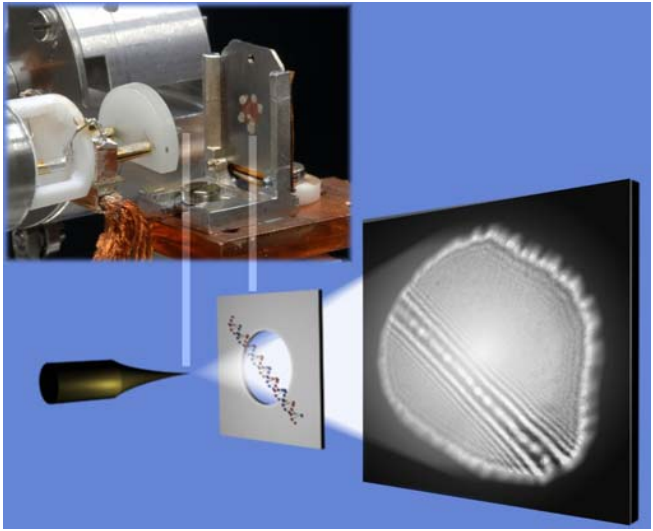

**FIG.1 Schematic of the set-up. For exploring radiation damage effects on DNA, the molecules are exposed to coherent low energy electrons. A spherical wave emitted from a coherent electron source is scattered at DNA molecules stretched over holes in a thin film and positioned at about 1 μm beyond the source. At a 10cm distant micro-channel-plate screen detector the interference between the elastic scattered wave (object wave) and the un-scattered wave (reference wave) produces the hologram captured by a 14bit dynamic range CCD camera. Magnification is provided by the geometry of the set-up alone and can be adjusted by the source-sample distance. The smallest interference fringe spacing in the hologram amounts to 0.7 nm and is a measure for the achievable resolution.**

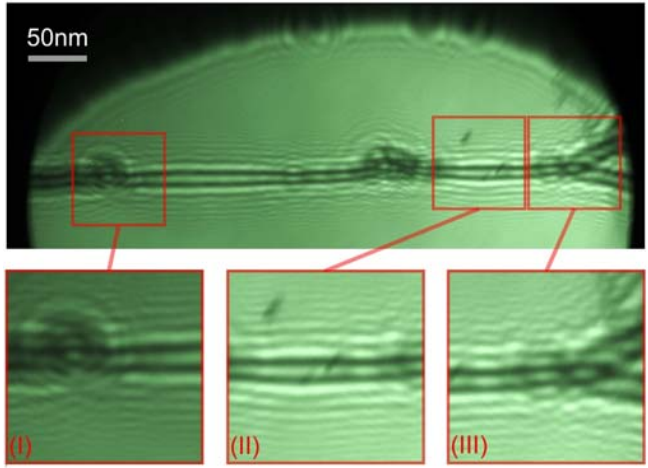
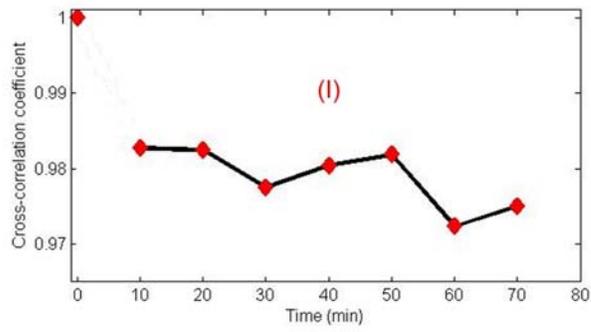
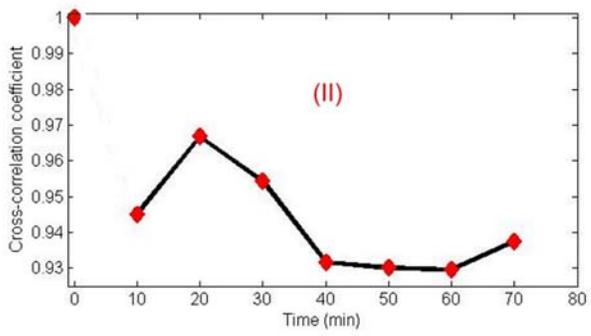
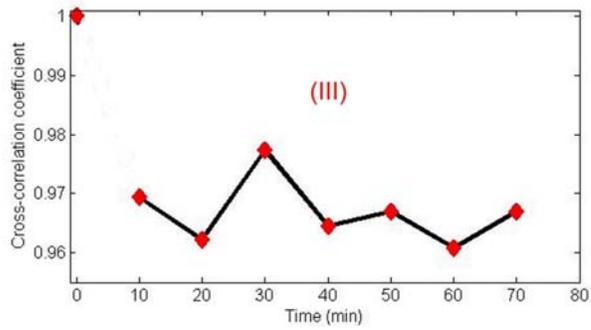

**FIG. 2 Non-destructive imaging of DNA.** The low energy electron hologram of DNA molecules stretched over a hole in a thin film imaged continuously for 70 minutes using electrons of 60eV kinetic energy and a total current of 200nA is shown on top. Part of the rim of the 1µm diameter hole, spanning the free standing molecules, is visible at the very top of the image. Three regions of the hologram, marked in red, have been chosen to evaluate the cross-correlation function of subsequent holograms. The evolution of the cross-correlation coefficient is shown in the diagrams corresponding to the three regions.

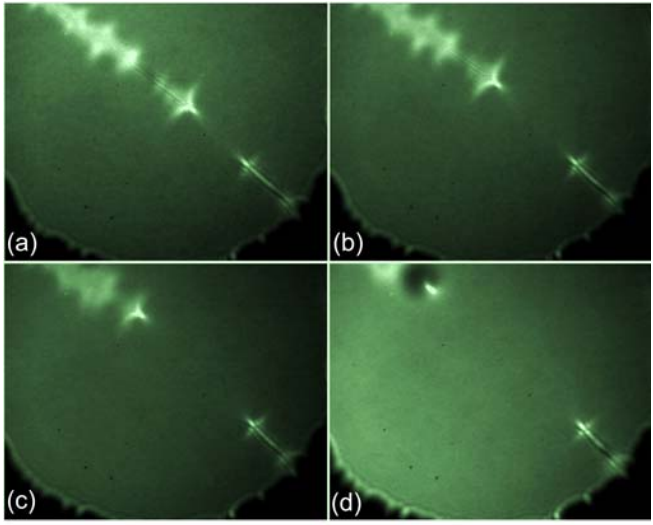
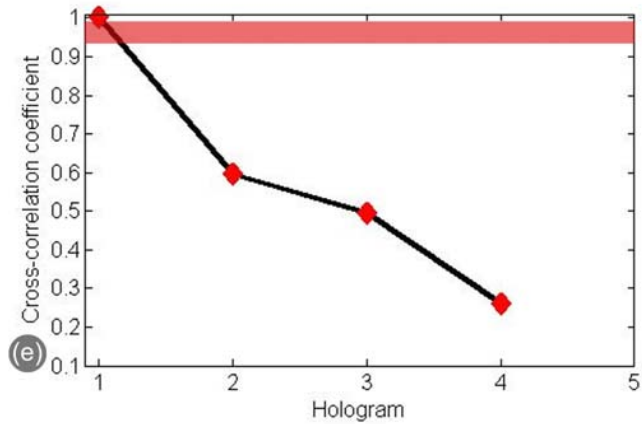

**FIG. 3 Decomposition of DNA. (a)-(d) Control experiment showing the rapid decomposition of DNA stretched over a 1μm diameter hole in a carbon film during imaging with 260eV electrons. At (e) the associated change of the cross-correlation coefficient is plotted. The red bar at the top indicates the range in which this coefficient fluctuated in the non-damaging experiment with 60eV electrons.**